\documentclass[aps,reprint,article,prx]{revtex4-1}
\usepackage{amsmath}
\usepackage{dcolumn}
\usepackage{graphicx}
\usepackage{float}
\usepackage{color}
\usepackage{ulem}

\begin{document}

\title{A generalized phenomenological model for the magnetic field penetration and magnetization hysteresis loops of a type-II superconductor}
\author{Wei Xie, Yu-Hao Liu, and Hai-Hu Wen$^{\dag}$}

\affiliation{National Laboratory of Solid State Microstructures and Department of Physics, Collaborative Innovation Center of Advanced Microstructures, Nanjing University, Nanjing 210093, China}

\begin{abstract}
A generalized phenomenological model for the critical state of type-II superconductors with magnetic field parallel to the superconducting plate is proposed. This model considers the global magnetization including both the equilibrium magnetization from surface screening current and the non-equilibrium magnetization from bulk pinning in a self-consistent way. Our model can be used to simulate the magnetization-hysteresis-loops (MHLs) and flux penetrating process of different type-II superconductors, from low- to high-$\kappa$ values. Here we take an optimally doped Ba$_{0.6}$K$_{0.4}$Fe$_2$As$_2$ single crystal as a testing example. The model can fit the data quite well and several important parameters can be extracted from the fitting. Thus, the model can be extended to a general case for studying the magnetization and flux penetration in other type-II superconductors.
\end{abstract}

\maketitle
\section{Introduction}

Due to different signs of the interface energy between superconducting and normal regions \cite{Abrikosov}, superconductors can be divided into two types: type-I superconductors with positive interface energy; type-II superconductors with negative interface energy. This categorization can be made based on the value of Ginzburg-Landau parameter $\kappa = \lambda/\xi$, for type-I superconductors $\kappa < 1/\sqrt{2}$, but for type-II superconductors $\kappa > 1/\sqrt{2}$. For a clean type-I superconductor, there is a well-defined thermodynamic critical field $H_\mathrm{c}$ which can be measured directly. When the applied field is below $H_\mathrm{c}$, the superconductor is in Meissner state. With the existence of surface screening supercurrent, flux can be fully expelled from the interior of superconductor. Above $H_\mathrm{c}$, the superconductor will lose its superconductivity and go into normal state. For a clean type-II superconductor, there are two different critical fields, namely the lower critical field $H_\mathrm{c1}$ and the upper critical field $H_\mathrm{c2}$. When the applied field is below $H_\mathrm{c1}$, the superconductor is in Meissner state, just as the type-I superconductor. From $H_\mathrm{c1}$ to $H_\mathrm{c2}$, the superconductor is in mixed state with vortices penetrating into the bulk region. Above $H_\mathrm{c2}$, the superconductor is in normal state \cite{Abrikosov,Ginzburg}. The determination of critical fields is important in the application of superconductors, such as the radio frequency superconducting cavities for the studies of high energy physics \cite{accelerator,cavity}. While in practice, there are defects, disorders and impurities in the superconductors, which can act as pinning centres of vortices \cite{Blatter,pinning,secondpeak}. In this case, in addition to the surface screening supercurrent, there exists a bulk supercurrent owing to these pinning centres. This critical current induced by pinning plays an important role in the application of superconductors \cite{criticalcurrent,Ishida,defects,transportcurrent}, and these superconductors with pinning centres are called hard superconductors. In the magnetization hysteresis loops (MHLs) of hard type-II superconductors, hysteresis can be easily seen.

In order to calculate the magnetization of hard type-II superconductors, Bean et. al. firstly proposed the critical state model \cite{BeanPRL,BeanRMP,currentloss} which can describe the flux penetration and vortex pinning, by assuming that: (1) the pinning force is uniform and independent of local magnetic field; (2) the Lorentz force is equal to the pinning force everywhere inside the superconductor. In the original model, the critical current density $J_\mathrm{c}$ calculated from magnetization is constant and independent of the applied field at a given temperature. This leads to a flat and parallel MHL both in the field ascending and descending processes. While in most experiments, it was found that the width of MHL curves will either decrease or increase with varying magnetic field \cite{flux,YBCO,vortexdynamics}. To figure out this problem, Kim et. al. assumed that the critical current density $J_\mathrm{c}$ due to bulk pinning is inversely proportional to the field, i.e., $J_\mathrm{c}(H_\mathrm{i}) = k/(A_1+|H_\mathrm{i}|)$ \cite{Kim1,Kim2,fluxcreep,fluxmotion}, here $H_\mathrm{i}$ is the local magnetic field. Afterwards, many other models were proposed for interpreting the field dependence of bulk pinning. For example, Watson et. al. considered the $J_\mathrm{c}$ to be linearly proportional to the field $J_\mathrm{c}(H_\mathrm{i}) = A_2+B_2|H_\mathrm{i}|$ \cite{Watson}; Irie et. al. proposed a power-law relationship $J_\mathrm{c}(H_\mathrm{i}) = k/(|H_\mathrm{i}|)^{n}$ \cite{Irie,Green}; Fitz et. al. introduced an exponential-law relationship $J_\mathrm{c}(H_\mathrm{i}) = J_\mathrm{c}(0)\exp(-|H_\mathrm{i}|/A_3)$ \cite{Fietz,Karasik,exponential}; Xu et. al. adopted a general form of $J_\mathrm{c}$ given by $J_\mathrm{c}(H_\mathrm{i}) = J_\mathrm{c}(0)/(1+|H_\mathrm{i}|/A_4)^{n}$ \cite{Xu}. In above equations, $A_1, A_2,B_2,k, A_3, A_4$ are fitting parameters.

Additionally, for the magnetization of low-$\kappa$ type-II superconductors, Kes et al. suggested to divide the magnetization into reversible part and irreversible part \cite{Kes}. Following this idea, Chen et al. also proposed an extended critical state model by combining the equilibrium magnetization and energy barrier of the surface layer with the non-equilibrium magnetization of the bulk region, the latter is described by the critical state model with modified boundary conditions \cite{surface,surfacebarrier,Sinder,Krasnov,Kimishima1,Kimishima2}. Furthermore, Gokhfeld et al. proposed a modified critical state model by assuming a field-dependent thickness of surface layer with screening current \cite{Gokhfeld1,Gokhfeld2}. When the magnetic field is reduced from a positive high value to less than $H_\mathrm{c1}$, there are still a lot of vortices inside the superconductor, while the surface screening layer looks like that in the initial penetration Meissner state, we call this state as vortex-trapped Meissner state. This has never been seriously dealt in previous models, only Matsushita et al. had a rough and macroscopic investigation on it and put forward another modified critical state model \cite{Matsushita}. Brandt et al. also considered the isotropy and anisotropy of superconductors with two different kinds of pinning in their $J_\mathrm{c}$ dependence for the critical state model and did some calculations on the global magnetization \cite{Mikitik}.

However, none of these models mentioned above can describe the magnetization of most type-II superconductors with different $\kappa$ values or ratios of $H_\mathrm{c2}/H_\mathrm{c1}$. We try to deal with this problem mathematically and take different kinds of type-II superconductors into consideration, and thus give a simple phenomenological model for the critical state in this paper. To validate our model, we measured the MHLs of optimally doped iron based superconductor Ba$_{0.6}$K$_{0.4}$Fe$_2$As$_2$ and found that the model fits experimental data in a self-consistent way. This paper is organized as follows. Section II introduces the basic assumptions of our model. Section III gives a brief description of the flux penetration process and calculation of magnetization. Section IV gives the result of experiments and numerical fittings. Section V gives a conclusion of our paper.

\section{Basic Assumptions of the model}

Our model also divides the magnetization into two major contributions, namely the equilibrium magnetization and the non-equilibrium magnetization, as adopted by Kes et al. and Chen et al. \cite{Kes,surface,surfacebarrier,Bi2212}. A complete MHL curve involves three major periods: the initial field penetrating process with the applied field $H_\mathrm{a}$ from zero to the maximum magnetic field $H_\mathrm{m}$, the second process with $H_\mathrm{a}$ from $H_\mathrm{m}$ to $-H_\mathrm{m}$, and the third process with $H_\mathrm{a}$ from $-H_\mathrm{m}$ to $H_\mathrm{m}$. In this section, we firstly handle with the equilibrium magnetization of type-II superconductors. Then, we deal with the non-equilibrium magnetization using a modified critical state model. Finally, we give a discussion on the vortex-trapped Meissner state both in the field descending and ascending process.

\subsection{Equilibrium Magnetization}
For the equilibrium magnetization, we consider a clean type-II superconductor without pinning centres inside the sample. It has a reversible magnetization curve $M(H)$ with two thermodynamic critical fields, namely $H_\mathrm{c1}$ and $H_\mathrm{c2}$. In order to calculate the magnetization, we separate the superconductor into two regions, the bulk region and a surface layer surrounding it \cite{surface}. The shielding current is flowing in the surface layer which is in the order of London penetration depth $\lambda_\mathrm{L}$. When the applied field is less than the lower critical field, i.e., $H_\mathrm{a} < H_\mathrm{c1}$, the superconductor is in Meissner state. The shielding supercurrent is flowing in the surface layer, and thus there exists a boundary between the surface layer and the inner bulk region. We define the field at the boundary $x = \lambda_\mathrm{L}$ as $H_\mathrm{e}$. In this case, the effective boundary field $H_\mathrm{e}$ should be zero, as well as the field in bulk region. When $H_\mathrm{a}$ is increased from $H_\mathrm{c1}$ to $H_\mathrm{c2}$, the superconductor is in mixed state and the shielding effect of surface supercurrent will be gradually reduced. Flux lines start to penetrate into the bulk region at $H_\mathrm{a} = H_\mathrm{c1}$. As there is no pinning centre, and thus no bulk supercurrent inside the superconductor, the field in the bulk region should be uniform and equal to $H_\mathrm{e}$. In the surface layer, the magnetic field decays exponentially in space from $H_\mathrm{a}$ to $H_\mathrm{e}$, in the scale of London penetration depth $\lambda_\mathrm{L}$ \cite{London}. When $H_\mathrm{a} > H_\mathrm{c2}$, the superconductor is in normal state and the field everywhere is equal to $H_\mathrm{a}$. Following these discussions, when the superconductor is in mixed state, we suppose a general relation between $H_\mathrm{e}$ and $H_\mathrm{a}$:
\begin{equation}\label{equ1}
M_\mathrm{equ} = -\frac{H_\mathrm{c1}(H_\mathrm{c2}-H_\mathrm{a})}{H_\mathrm{c2}-H_\mathrm{c1}}(\frac{H_\mathrm{c1}}{H_\mathrm{a}})^{\alpha}.
\end{equation}
\begin{equation}\label{equ2}
H_\mathrm{e} = M_\mathrm{equ}+H_\mathrm{a}.
\end{equation}
Here $H_\mathrm{c1}$ and $H_\mathrm{c2}$ are the lower critical field and the upper critical field, respectively. $\alpha$ is a dimensionless fitting parameter. $M_\mathrm{equ}$ is the equilibrium magnetization. This relation satisfies the boundary conditions of the mixed state: when $H_\mathrm{a} = H_\mathrm{c1}$, the mixed state has $M_\mathrm{equ} = -H_\mathrm{c1}$, $H_\mathrm{e} = 0$, and when $H_\mathrm{a} = H_\mathrm{c2}$, the mixed state has $M_\mathrm{equ} = 0$, $H_\mathrm{e} = H_\mathrm{c2}$.  The derivative of $M_\mathrm{equ}$ with respect to $H_\mathrm{a}$ is the magnetic susceptibility of the equilibrium magnetization curve which is defined as $\chi_\mathrm{equ}$:
\begin{equation}\label{equ3}
\chi_\mathrm{equ} = \frac{\mathrm{d}M_\mathrm{equ}}{\mathrm{d}H_\mathrm{a}} = \frac{H_\mathrm{c1}}{H_\mathrm{c2}-H_\mathrm{c1}}(\frac{H_\mathrm{c1}}{H_\mathrm{a}})^{\alpha}[1+\frac{\alpha(H_\mathrm{c2}-H_\mathrm{a})}{H_\mathrm{a}}].
\end{equation}
When $H_\mathrm{a} = H_\mathrm{c2}$, we have
\begin{equation}\label{equ4}
\chi_\mathrm{equ,c2} = \frac{H_\mathrm{c1}}{H_\mathrm{c2}-H_\mathrm{c1}}(\frac{H_\mathrm{c1}}{H_\mathrm{c2}})^{\alpha}.
\end{equation}
In this way, the fitting parameter $\alpha$ can be written as
\begin{equation}\label{equ5}
\alpha = \ln(\frac{(H_\mathrm{c2}-H_\mathrm{c1})\chi_\mathrm{equ,c2}}{H_\mathrm{c1}})/\ln(\frac{H_\mathrm{c1}}{H_\mathrm{c2}})
\end{equation}
If $H_\mathrm{c2}$ is much larger than $H_\mathrm{c1}$ ($H_\mathrm{c1} \ll H_\mathrm{c2}$, especially for high-$\kappa$ superconductors, such as cuprate and iron-based superconductors), the equation can be simplified to
\begin{equation}\label{equ6}
\alpha = -1+\ln(\chi_\mathrm{equ,c2})/\ln(\frac{H_\mathrm{c1}}{H_\mathrm{c2}}).
\end{equation}
The $H_\mathrm{c2}$ is always tens or hundreds of Tesla and difficult to be achieved in experiment. Thus, we may also pay some attention to the magnetic susceptibility at the lower critical field. When $H_\mathrm{a} = H_\mathrm{c1}$, we have
\begin{equation}\label{equ7}
\chi_\mathrm{equ,c1} = \frac{H_\mathrm{c1}}{H_\mathrm{c2}-H_\mathrm{c1}}+\alpha.
\end{equation}
The fitting parameter $\alpha$ can be written as
\begin{equation}\label{equ8}
\alpha = \chi_\mathrm{equ,c1}-\frac{H_\mathrm{c1}}{H_\mathrm{c2}-H_\mathrm{c1}}.
\end{equation}
If $H_\mathrm{c1} \ll H_\mathrm{c2}$, the fitting parameter $\alpha$ is right the magnitude of the magnetic susceptibility at $H_\mathrm{c1}$. Above all, if we know three of the four variables $\chi_\mathrm{equ,c1}$ or $\chi_\mathrm{equ,c2}$, $H_\mathrm{c1}$, $H_\mathrm{c2}$, $\alpha$, the last one can therefore be obtained.

As a comparison, the reversible relation of the mixed state adopted by Kes et al. fit experimental data only in the high temperature region for low-$\kappa$ type-II superconductors \cite{Kes}. Chen et al. used an exponential relation of $H_\mathrm{e}$ and $H_\mathrm{a}$ when the maximum magnetic field $H_\mathrm{m}$ is much lower than $H_\mathrm{c2}$ \cite{surface}. But when $H_\mathrm{a}$ is close enough to $H_\mathrm{c2}$ or the $\kappa$ is high, the relation $H_\mathrm{e}(H_\mathrm{a})$ is inclined to be a straight line \cite{Hao1,Hao2}. Our relation combines the merits of both ideas above. In the low field limit ($H_\mathrm{a}$ is close to $H_\mathrm{c1}$), if $H_\mathrm{c1} \ll H_\mathrm{c2}$, the relation can be simplified to a power law:
\begin{equation}\label{equ9}
M_\mathrm{equ} \propto -H_\mathrm{a}^{-\alpha},
\end{equation}
with $(H_\mathrm{c2}-H_\mathrm{a})/(H_\mathrm{c2}-H_\mathrm{c1})\rightarrow 1$. However, this case ($H_\mathrm{c1} \ll H_\mathrm{c2}$) was not considered in the model of Kes, in which the relation $B(H_\mathrm{a})$ was simplified to a linear one and is not consistent with the experiment \cite{Kes}. In the high field limit ($H_\mathrm{a}$ is close to $H_\mathrm{c2}$), if $H_\mathrm{c1} \ll H_\mathrm{c2}$, the $M_\mathrm{equ}$ in Eq.~\ref{equ1} is small compared with $H_\mathrm{a}$ and our relation $H_\mathrm{e}(H_\mathrm{a})$ is approximately to be a straight line, which is consistent with that suggested by Campbell et al.\cite{campbell}.

For simplicity, we assume the spatial distribution of field in the surface layer to be linear. This can serve as a good approximation when the sample dimension is much larger than the London penetration depth $\lambda_\mathrm{L}$. Furthermore, the London penetration depth $\lambda_\mathrm{L}$ is related to the measuring temperature and this will lead to a more complicated relation \cite{Gokhfeld1,Gokhfeld2}. In our model, we assume $\lambda_\mathrm{L}$ to remain unchanged with magnetic field for simplicity. The fitting will be done for magnetization hysteresis loop measured at a particular temperature, thus the penetration depth $\lambda_\mathrm{L}$ will be a fitting parameter for each temperature.

\subsection{Non-equilibrium Magnetization}

For the non-equilibrium magnetization, it is a little different because of the existence of pinning centres. For a hard type-II superconductor with pinning centers, the magnetization hysteresis $M(H)$ is irreversible. Under our assumptions, the shielding effect of surface layer is just the same as the equilibrium part introduced above. The boundary field $H_\mathrm{e}$ is self-consistently determined by the same value at the boundary, which connects the surface screening layer and the bulk region, as Eq.~\ref{equ1} and Eq.~\ref{equ2}. While for the bulk region, the flux motion should be prevented due to the existence of pinning centers. Bean et al. \cite{BeanPRL,BeanRMP} proposed a critical state, in which the pinning force and the magnetic pressure (or Lorenz force) originating from the gradient of magnetic field are in equilibrium. Thus, the field in the bulk region will not be uniform anymore. In this case, the superconductor can carry a critical current which obeys the law $\nabla\times B = \mu_\mathrm{0}J_\mathrm{c}$, with $B$ the local average density of the magnetic induction. In the critical state model, this pinning force is equal to Lorenz force $\Phi_{0}J_\mathrm{0}$. There are different forms of $J_\mathrm{c}$ with respect to magnetic field proposed previously, as we have discussed in section I. But most of these models did not concern the upper critical field $H_\mathrm{c2}$ in their $J_\mathrm{c}$ equations. In fact, the critical current density $J_\mathrm{c}$ should be zero when the applied field $H_\mathrm{a}$ approaches $H_\mathrm{c2}$. To deal with this problem, we assume a more general form of $J_\mathrm{c}$ in our model, this gives not only a self-consistent formula of the boundary field $H_\mathrm{e}$, but also on the lower critical field $H_\mathrm{c1}$ and the upper critical field $H_\mathrm{c2}$, as shown by the following equation
\begin{equation}\label{equ10}
J_\mathrm{c} = J_\mathrm{c0}\frac{H_\mathrm{c2}-|H_\mathrm{e}|}{H_\mathrm{c2}}(\frac{H_\mathrm{c1}}{H_\mathrm{c1}+|H_\mathrm{e}|})^{\beta}.
\end{equation}
Here $H_\mathrm{c1}$ and $H_\mathrm{c2}$ are the lower critical field and the upper critical field, respectively. $\beta$ is a dimensionless fitting parameter. $J_\mathrm{c0}$ is the critical current density when the field is zero. This relation satisfies the experimental variation of critical current density very well. When $H_\mathrm{e}$ = 0, the critical current density is equal to $J_\mathrm{c0}$. When $H_\mathrm{e} = H_\mathrm{c2}$, the superconductor is in normal state and the critical current density is zero. In the low field limit ($H_\mathrm{a}$ is close to $H_\mathrm{c1}$), if $H_\mathrm{c1} \ll H_\mathrm{c2}$, the equation can be simplified to a power law:
\begin{equation}\label{equ11}
J_\mathrm{c} \propto (H_\mathrm{c1}+|H_\mathrm{e}|)^{-\beta}
\end{equation}
with $(H_\mathrm{c2}-|H_\mathrm{e}|)/H_\mathrm{c2} \rightarrow 1$. When $\beta$ = 1, the equation is right the relation adopted by Kim \cite{Kim1}. In the high field limit ($H_\mathrm{a}$ is close to $H_\mathrm{c2}$), the equation tends to be a straight line, which is consistent with the experiment \cite{Livingston}. Above all, our equation treats the magnetic induction and critical current density in a more reasonable way. With the Maxwell equation for the one-dimensional superconductor $-\mathrm{d}H(x)/\mathrm{d}x = J_\mathrm{c}$ and proper boundary conditions, we can calculate the spatial distribution of magnetic induction inside the superconductor.

\subsection{Vortex-trapped Meissner State}

In the field descending process, when the applied field $H_\mathrm{a}$ is decreased below $H_\mathrm{c1}$, the superconductor will try to re-enter the Meissner state. But this Meissner state is different from that of the initial field penetrating process, as the bulk pinning is different in these two cases. In the initial Meissner state, no vortex penetrate into the bulk region, and thus, there is no bulk pinning current inside the superconductor. But in the Meissner state of the field descending process, vortices will be trapped inside the superconductor and the bulk pinning current is not zero anymore. This remaining bulk pinning current is also different from the pinning current $J_\mathrm{c}$ given by Eq.~\ref{equ10} as $H_\mathrm{e} = 0$, and we defined it as $J_\mathrm{cr}$. From the experimental perspective, we can see a sudden change of the slope of the linear part from magnetization curve when $H_\mathrm{a} < H_\mathrm{c1}$ in these two cases. This is particularly obvious when the Ginzburg$-$Landau parameter $\kappa$ is small or $H_\mathrm{c1}$ is in the same order of magnitude with $H_\mathrm{c2}$ \cite{Matsushita}. This sudden change was not explained in the models of Chen \cite{surface} and Walmsley \cite{Walmsley}. In their models, they assumed the bulk pinning in the superconductor to remain unchanged when $H_\mathrm{a} < H_\mathrm{c1}$ in the field descending process, which is clearly unreasonable. Because the flux distribution profile in the bulk region during this period varies remarkably, the magnetization arising from the bulk pinning will change along with the varying applied field. Matsushita et al.\cite{Matsushita} dealt this problem and offered a phenomenological model with a perfect Meissner state layer and a flux trapped interior part, and the boundary of both regions will move towards the interior region of the bulk with descending field.

\begin{figure*}[htbp]
\centering
\includegraphics[width=11cm]{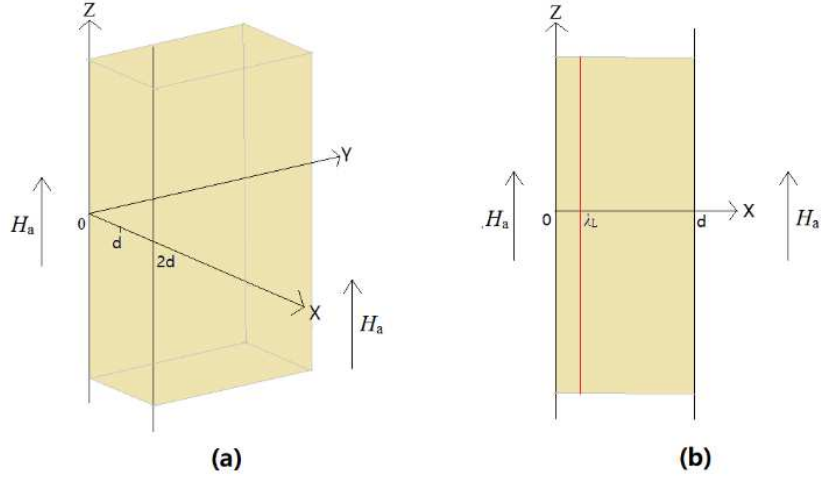}
\caption{Sketch of an infinite large superconducting plate: (a) The superconducting plate with thickness of $2d$ along $x$-axis. The field is applied along $z$-axis. (b) The cross-section of half plate in (a) along $x$-axis. The light blue region indicates the superconducting plate. The red line shows the boundary of surface layer and bulk region. The London penetration depth $\lambda_\mathrm{L}$ is much less than the thickness of plate $2d$ ($\lambda_\mathrm{L} << 2d$). Here we have magnified $\lambda_\mathrm{L}$ to make the surface layer discernible.
} \label{fig1}
\end{figure*}

Unlike these models above, we propose a new model to deal with this sophisticated situation, based on the ideas of critical state. In the field descending process, when the applied field $H_\mathrm{a}$ is decreased below $H_\mathrm{c1}$, we assume the shielding supercurrent to be the same as the initial Meissner state. Thus, the surface layer will be in full shielding state and the boundary field $H_\mathrm{e}$ is zero as before. For a vortex existing near the boundary of bulk region and surface layer, it feels Lorenz forces both from the shielding supercurrent $J_\mathrm{s}$ and the remaining bulk pinning current $J_\mathrm{cr}$ with opposite directions. These two forces, as well as the pinning force $F_\mathrm{p}$ from disorders and defects, form an equilibrium state $\Phi_{0}J_\mathrm{s} - \Phi_{0}J_\mathrm{cr} + F_\mathrm{p} = 0$. When $H_\mathrm{a}$ is further decreased, the shielding supercurrent $J_\mathrm{s}$, which is the gradient of magnetic field in the surface layer, will decrease simultaneously as a linear approximation $J_\mathrm{s} \propto (H_\mathrm{a}-H_\mathrm{e})/\lambda_\mathrm{L} = H_\mathrm{a}/\lambda_\mathrm{L}$. Since $J_\mathrm{s}$ is related to the force preventing the trapped vortices from out-going, the remaining bulk pinning current $J_\mathrm{cr}$ in the bulk region will reduce in the same pace with the applied field $H_\mathrm{a}$ to form a new equilibrium state. Above all, we assume that $J_\mathrm{cr}$ varies linearly with $H_\mathrm{a}$:
\begin{equation}\label{equ12}
J_\mathrm{cr}=J_\mathrm{c0}\frac{H_\mathrm{a}-H_\mathrm{0}}{H_\mathrm{c1}-H_\mathrm{0}}
\end{equation}
$J_\mathrm{c0}$ is the zero-field critical current density from equation, $H_\mathrm{c1}$ is the lower critical field, $H_\mathrm{0}$ is the field at which the remaining pinning current density reduces to zero. In this case, the magnetization of Meissner state in our model is linearly varying with applied field either in the initial process and the field descending process. In next section, we will prove in detail that, $H_\mathrm{0}$ can be simply obtained from fitting to the experimental data, which is the intersection point of extended lines of magnetization curves from both the initial Meissner state and the field descending Meissner state.

The description of magnetization in other regions, for example from $H_a=$ $-H_{c1}$ to $-H_m$ (with $H_m\gg H_{c1}$) and from $-H_m$ to zero will be the replica of the status mentioned above.

\section{Calculation of Magnetization in Different Regions}

After establishing the model as mentioned in section II, the magnetization in different regions of applying magnetic field can be obtained. The superconductor in our model is assumed to be a plate with infinite lateral sizes and thickness of $2d$ along $x$-axis, as shown in Fig.~\ref{fig1}(a). The magnetic field $H_\mathrm{a}$ is applied parallel to the plate and enters from both sides of it. For convenience, we set the field direction along $z$-axis, and as a consequence, the demagnetization factor is close to zero \cite{demag}. Due to the symmetry of the plate, the distribution of field or current is one-dimensional along $x$-axis. Thus, we take the cross-section of this plate along $x$-axis, as shown in Fig.~\ref{fig1}(b). In Fig.~\ref{fig1}(b), $x = 0$ represents the interface of superconductor and vacuum, $\lambda_\mathrm{L}$ is the London penetration depth which is much smaller than the sample dimension ($\lambda_\mathrm{L} << 2d$), $x = d$ is the centre of the plate. We have only shown half of the plate in Fig.~\ref{fig1}(b) and another half can be obtained directly with mirror symmetry of $x = d$.

A complete field penetrating process includes three parts: the initial field penetrating process, the field ascending process and the field descending process. Before magnetization measurement, the superconductor should be zero-field cooled down to a certain temperature, which is below the critical temperature $T_\mathrm{c}$. The field distribution in the surface layer is $H_\mathrm{s}(x)$ (with $0 < x < \lambda_\mathrm{L}$) and in the bulk region is $H_\mathrm{b}(x)$ (with $\lambda_\mathrm{L} < x < d$). For simplicity, the field distributions of both surface layer and bulk region are linear in space. If $H_\mathrm{c1} < H_\mathrm{m} < H_\mathrm{c2}$, details of the three parts are as follows.

1. The initial field penetrating process

\begin{figure}[htbp]
\centering
\includegraphics[width=8cm]{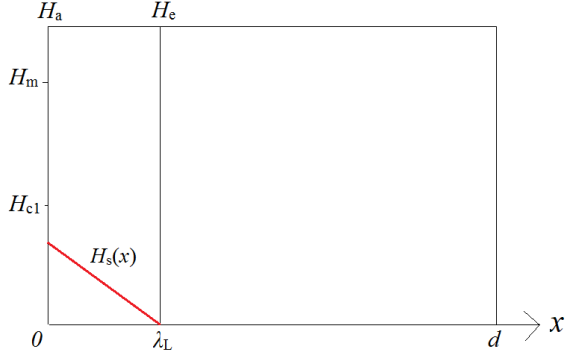}
\caption{Magnetic field distribution (red line) of a superconductor in Stage ($\mathbf{i}$) (when $H_\mathrm{a} < H_\mathrm{c1}$). $H_\mathrm{s}(x)$ is the distribution function in the surface layer.
} \label{fig2}
\end{figure}

Stage ($\mathbf{i}$): $H_\mathrm{a}$ is increased from 0 to $H_\mathrm{c1}$. The superconductor is in Meissner state. The magnetic field is confined in the surface layer which is in the order of $\lambda_\mathrm{L}$, and $H_\mathrm{e}$ is zero. Vortices have not penetrated the bulk region yet. Thus, there is only the equilibrium magnetization. For this stage, we have the field distribution as follows (as shown in Fig.~\ref{fig2})
\begin{equation}\label{equ31}
H_\mathrm{s}(x)=\frac{H_\mathrm{a}}{\lambda_\mathrm{L}}(\lambda_\mathrm{L}-x),
\end{equation}
\begin{equation}\label{equ32}
H_\mathrm{b}(x)=0.
\end{equation}

\begin{figure}[htbp]
\centering
\includegraphics[width=8cm]{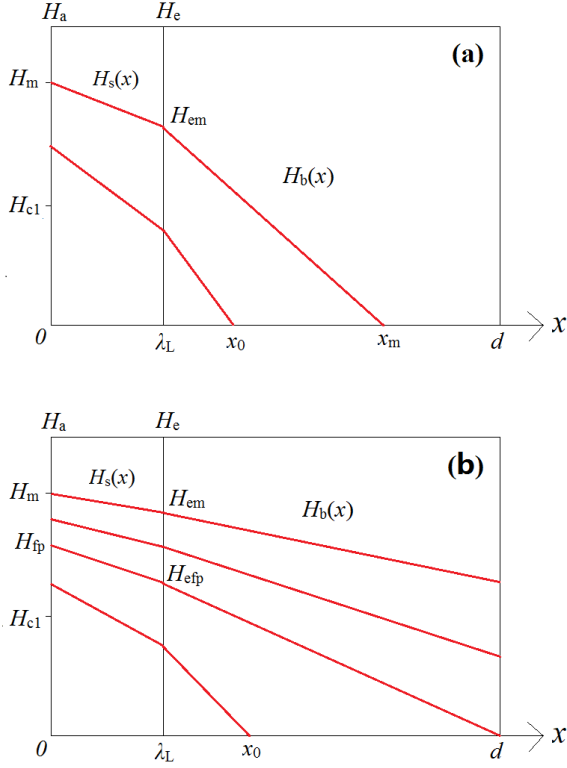}
\caption{Magnetic field distribution (red lines) of a superconductor in Stage ($\mathbf{ii}$) (when $H_\mathrm{c1} < H_\mathrm{a} < H_\mathrm{m}$) if: (a) $H_\mathrm{m} < H_\mathrm{fp}$; (b) $H_\mathrm{m} > H_\mathrm{fp}$. $H_\mathrm{s}(x)$ is the distribution function in the surface layer. $H_\mathrm{b}(x)$ is the distribution function in the bulk region.
} \label{fig3}
\end{figure}

Stage ($\mathbf{ii}$): $H_\mathrm{a}$ is increased from $H_\mathrm{c1}$ to $H_\mathrm{m} (< H_\mathrm{c2})$. The superconductor is in mixed state and the shielding effect of surface layer will be partially reduced. And thus, vortices start to penetrate into the bulk region. There are both equilibrium magnetization and non-equilibrium magnetization. We define the field where the sample is fully penetrated by the vortices as $H_\mathrm{fp}$. There are two cases in this stage: $H_\mathrm{m} < H_\mathrm{fp}$ and $H_\mathrm{m} > H_\mathrm{fp}$. In the former case, we have the field distribution as follows (as shown in Fig.~\ref{fig3}(a))
\begin{equation}\label{equ33}
H_\mathrm{s}(x)=-\frac{x}{\lambda_\mathrm{L}}(H_\mathrm{a}-H_\mathrm{e})+H_\mathrm{a}.
\end{equation}
\begin{equation}\label{equ34}
H_\mathrm{b}(x)=
\begin{cases}
-J_\mathrm{c}(x-\lambda_\mathrm{L})+H_\mathrm{e}& \text{$\lambda_\mathrm{L}<x<x_\mathrm{0}$}\\
0& \text{$x_\mathrm{0}<x<d$}
\end{cases}
\end{equation}
In the above equations, $H_\mathrm{e}$ is the boundary field between the surface layer and the bulk region, which is connected with the applied field $H_\mathrm{a}$ by Eq.~\ref{equ1} and Eq.~\ref{equ2}. $J_\mathrm{c}$ is the bulk pinning current density, which is connected with $H_\mathrm{e}$ by Eq.~\ref{equ10}. $x_\mathrm{0}$ is the frontier of the vortices, which is
\begin{equation}\label{equ35}
x_\mathrm{0}=\frac{H_\mathrm{e}}{J_\mathrm{c}}+\lambda_\mathrm{L}.
\end{equation}
The fully penetration field $H_\mathrm{fp}$ can be obtained in the following way. When $H_\mathrm{a} = H_\mathrm{fp}$, the boundary field is $H_\mathrm{efp} = H_\mathrm{e}[H_\mathrm{fp}]$ and the flux front reaches right at the center of the superconductor $x_\mathrm{0} = d$. Thus we have
\begin{equation}\label{equ36}
H_\mathrm{efp}=J_\mathrm{c}(d-\lambda_\mathrm{L}).
\end{equation}
We calculate Eq.~\ref{equ36} and Eq.~\ref{equ10} to get $H_\mathrm{efp}$, and the fully penetration field $H_\mathrm{fp}$ can therefore be obtained from Eq.~\ref{equ1} and Eq.~\ref{equ2}.

If $H_\mathrm{m} > H_\mathrm{fp}$, the situation is somewhat complicated. When $H_\mathrm{a} < H_\mathrm{fp}$ (or the flux front position $x_0$ is less than the sample thickness $x_\mathrm{0} < d$), the field distribution functions are just described by Eq.~\ref{equ33} and Eq.~\ref{equ34}. When $H_\mathrm{a} > H_\mathrm{fp}$ (or $x_\mathrm{0} > d$), the superconductor has been fully penetrated by vortices. The field distribution $H_\mathrm{s}(x)$ is described by Eq.~\ref{equ33} and the $H_\mathrm{b}(x)$ is given by (as shown in Fig.~\ref{fig3}(b))
\begin{equation}\label{equ38}
H_\mathrm{b}(x)=-J_\mathrm{c}(x-\lambda_\mathrm{L})+H_\mathrm{e}.
\end{equation}

2. The field descending process

\begin{figure}[htbp]
\centering
\includegraphics[width=8cm]{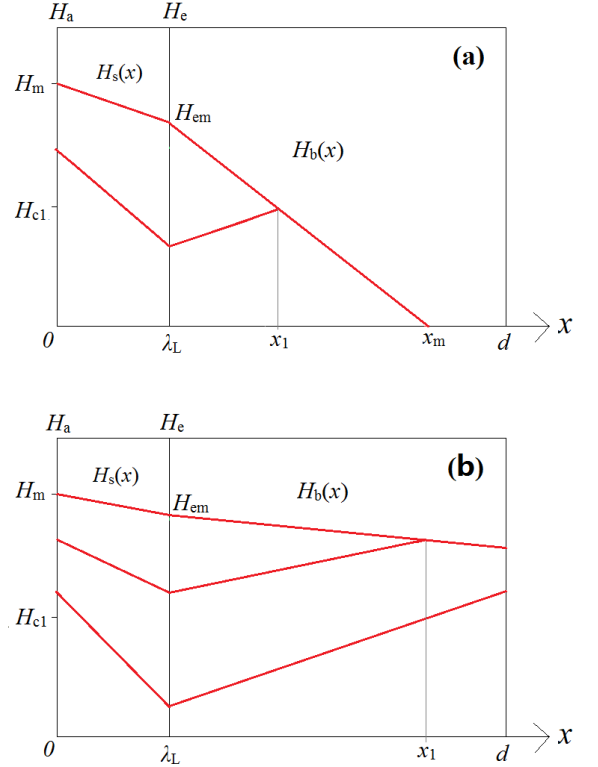}
\caption{Magnetic field distribution (red lines) of a superconductor in Stage ($\mathbf{iii}$) (when $H_\mathrm{c1} < H_\mathrm{a} < H_\mathrm{m}$) if: (a) $H_\mathrm{m} < H_\mathrm{fp}$; (b) $H_\mathrm{m} > H_\mathrm{fp}$. $H_\mathrm{s}(x)$ is the distribution function in the surface layer. $H_\mathrm{b}(x)$ is the distribution function in the bulk region.
} \label{fig4}
\end{figure}

Stage ($\mathbf{iii}$): $H_\mathrm{a}$ is decreased from $H_\mathrm{m}$ to $H_\mathrm{c1}$. The superconductor is still in mixed state. The bulk pinning current will change its direction from the outer part of bulk region, while the inner part remains unchanged. We define the position where the bulk pinning current changes its direction as $x_\mathrm{1}$. The field distribution is continuous at this crossover. This stage can also be divided into two cases: $H_\mathrm{m} < H_\mathrm{fp}$ and $H_\mathrm{m} > H_\mathrm{fp}$. We define the maximum boundary field as $H_\mathrm{em} = H_\mathrm{a}[H_\mathrm{m}]$ and the corresponding critical current density is $J_\mathrm{em} = J_\mathrm{c}[H_\mathrm{em}]$. If $H_\mathrm{m} < H_\mathrm{fp}$, the flux will not fully penetrate the superconductor and the maximum penetration depth is defined as $x_\mathrm{m}$
\begin{equation}\label{equ39}
x_\mathrm{m}=\frac{H_\mathrm{em}}{J_\mathrm{cm}}+\lambda_\mathrm{L}.
\end{equation}
The position $x_\mathrm{1}$ can be obtained by
\begin{equation}\label{equ40}
x_\mathrm{1}=\frac{H_\mathrm{em}-H_\mathrm{e}}{J_\mathrm{cm}+J_\mathrm{c}}+\lambda_\mathrm{L}.
\end{equation}
In this way, $H_\mathrm{s}(x)$ is described by Eq.~\ref{equ33} and $H_\mathrm{b}(x)$ is as follows (as shown in Fig.~\ref{fig4}(a))
\begin{equation}\label{equ42}
H_\mathrm{b}(x)=
\begin{cases}
J_\mathrm{c}(x-\lambda_\mathrm{L})+H_\mathrm{e}& \text{$\lambda_\mathrm{L}<x<x_\mathrm{1}$}\\
-J_\mathrm{cm}(x-\lambda_\mathrm{L})+H_\mathrm{em}& \text{$x_\mathrm{1}<x<x_\mathrm{m}$}\\
0& \text{$x_\mathrm{m}<x<d$}
\end{cases}
\end{equation}
If $H_\mathrm{m} > H_\mathrm{fp}$, the position where the critical current changes its direction $x_\mathrm{1}$ is also obtained by Eq.~\ref{equ40}. With decreasing the applied field, the opposite direction current will gradually penetrate the superconductor and the dividing line $x = x_\mathrm{1}$ is approaching the center of superconductor $x = d$. In this case, the field distribution $H_\mathrm{s}(x)$ is described by Eq.~\ref{equ33}. When $x_\mathrm{1} < d$, $H_\mathrm{b}(x)$ is (as shown in Fig.~\ref{fig4}(b))
\begin{equation}\label{equ44}
H_\mathrm{b}(x)=
\begin{cases}
J_\mathrm{c}(x-\lambda_\mathrm{L})+H_\mathrm{e}& \text{$\lambda_\mathrm{L}<x<x_\mathrm{1}$}\\
-J_\mathrm{cm}(x-\lambda_\mathrm{L})+H_\mathrm{em}& \text{$x_\mathrm{1}<x<d$}
\end{cases}
\end{equation}
But when $x_\mathrm{1} > d$, $H_\mathrm{b}(x)$ is given by (as shown in Fig.~\ref{fig4}(b))
\begin{equation}\label{equ46}
H_\mathrm{b}(x)=J_\mathrm{c}(x-\lambda_\mathrm{L})+H_\mathrm{e}.
\end{equation}

\begin{figure}[htbp]
\centering
\includegraphics[width=8cm]{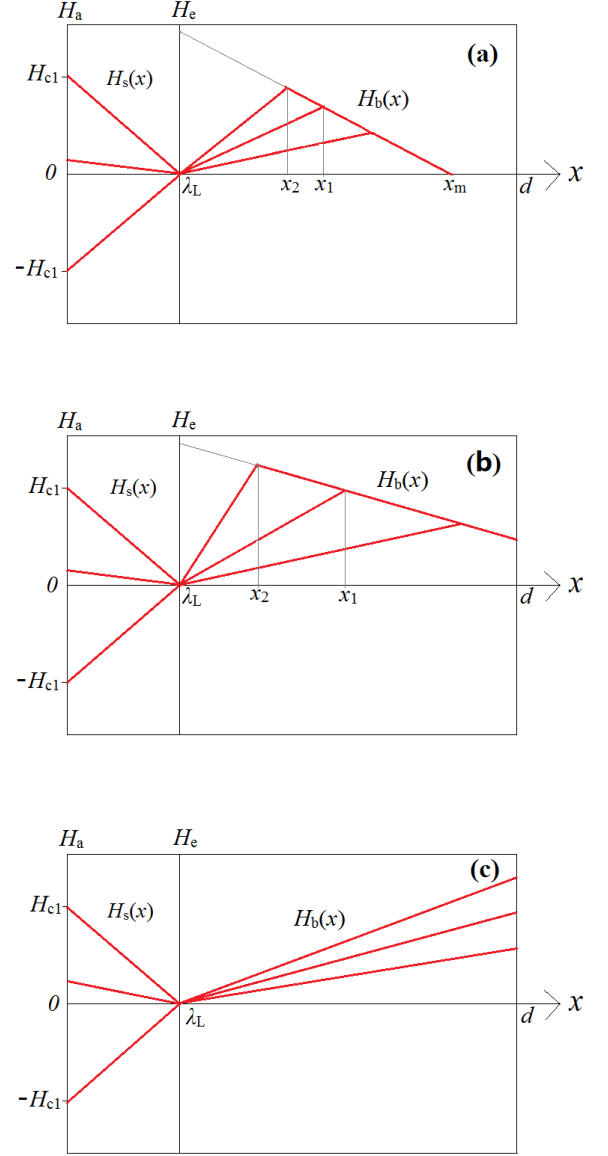}
\caption{Magnetic field distribution (red lines) of a superconductor in Stage ($\mathbf{iv}$) (when $-H_\mathrm{c1} < H_\mathrm{a} < H_\mathrm{c1}$) if: (a) $H_\mathrm{m} < H_\mathrm{fp}$; (b) $H_\mathrm{m} > H_\mathrm{fp}$ and $x_\mathrm{2} < d$; (c) $H_\mathrm{m} > H_\mathrm{fp}$ and $x_\mathrm{2} > d$. $H_\mathrm{s}(x)$ is the distribution function in the surface layer. $H_\mathrm{b}(x)$ is the distribution function in the bulk region.
} \label{fig5}
\end{figure}

Stage ($\mathbf{iv}$): $H_\mathrm{a}$ is decreased from $H_\mathrm{c1}$ to 0 and increased from 0 to $-H_\mathrm{c1}$. The superconductor re-enters Meissner state, which is different from the Meissner state in the initial penetrating process. Vortices are trapped in the bulk region, and thus, the non-equilibrium magnetization is not zero as Stage ($\mathbf{i}$). When $H_\mathrm{a}$ is decreased, the trapped vortices will go outwards and the remaining bulk pinning current density $J_\mathrm{cr}$ will change in the meantime, as described by Eq.~\ref{equ12}. We also divide this stage into two cases: $H_\mathrm{m} < H_\mathrm{fp}$ and $H_\mathrm{m} > H_\mathrm{fp}$. When $H_\mathrm{a} = H_\mathrm{c1}$, the initial dividing line $x_\mathrm{1}$ (given by Eq.~\ref{equ40}) is defined as $x_\mathrm{2}$. If $H_\mathrm{m} < H_\mathrm{fp}$, there are critical currents with different directions in the bulk region and the new dividing line $x_\mathrm{1}$ is given by
\begin{equation}\label{equ47}
x_\mathrm{1}=\frac{H_\mathrm{em}}{J_\mathrm{cm}+J_\mathrm{cr}}+\lambda_\mathrm{L}.
\end{equation}
The field distribution in the superconductor is (as shown in Fig.~\ref{fig5}(a))
\begin{equation}\label{equ48}
H_\mathrm{s}(x)=\frac{H_\mathrm{a}}{\lambda_\mathrm{L}}(\lambda_\mathrm{L}-x).
\end{equation}
\begin{equation}\label{equ49}
H_\mathrm{b}(x)=
\begin{cases}
J_\mathrm{cr}(x-\lambda_\mathrm{L})& \text{$\lambda_\mathrm{L}<x<x_\mathrm{1}$}\\
-J_\mathrm{cm}(x-\lambda_\mathrm{L})+H_\mathrm{em}& \text{$x_\mathrm{1}<x<x_\mathrm{m}$}\\
0& \text{$x_\mathrm{m}<x<d$}
\end{cases}
\end{equation}
If $H_\mathrm{m} > H_\mathrm{fp}$ and $x_\mathrm{2} < d$, there are critical currents with different directions in the bulk region and the new dividing line $x_\mathrm{1}$ is also given by Eq.~\ref{equ47}. $H_\mathrm{s}(x)$ is described by Eq.~\ref{equ48}. $H_\mathrm{b}(x)$ is (as shown in Fig.~\ref{fig5}(b))
\begin{equation}\label{equ51}
H_\mathrm{b}(x)=
\begin{cases}
J_\mathrm{cr}(x-\lambda_\mathrm{L})& \text{$\lambda_\mathrm{L}<x<x_\mathrm{1}$}\\
-J_\mathrm{cm}(x-\lambda_\mathrm{L})+H_\mathrm{em}& \text{$x_\mathrm{1}<x<d$}
\end{cases}
\end{equation}
If $H_\mathrm{m} > H_\mathrm{fp}$ and $x_\mathrm{2} > d$, the critical current has only one direction in the bulk region. $H_\mathrm{s}(x)$ is described by Eq.~\ref{equ48}. $H_\mathrm{b}(x)$ is (as shown in Fig.~\ref{fig5}(c))
\begin{equation}\label{equ53}
H_\mathrm{b}(x)=J_\mathrm{cr}(x-\lambda_\mathrm{L})
\end{equation}

\begin{figure}[htbp]
\centering
\includegraphics[width=8cm]{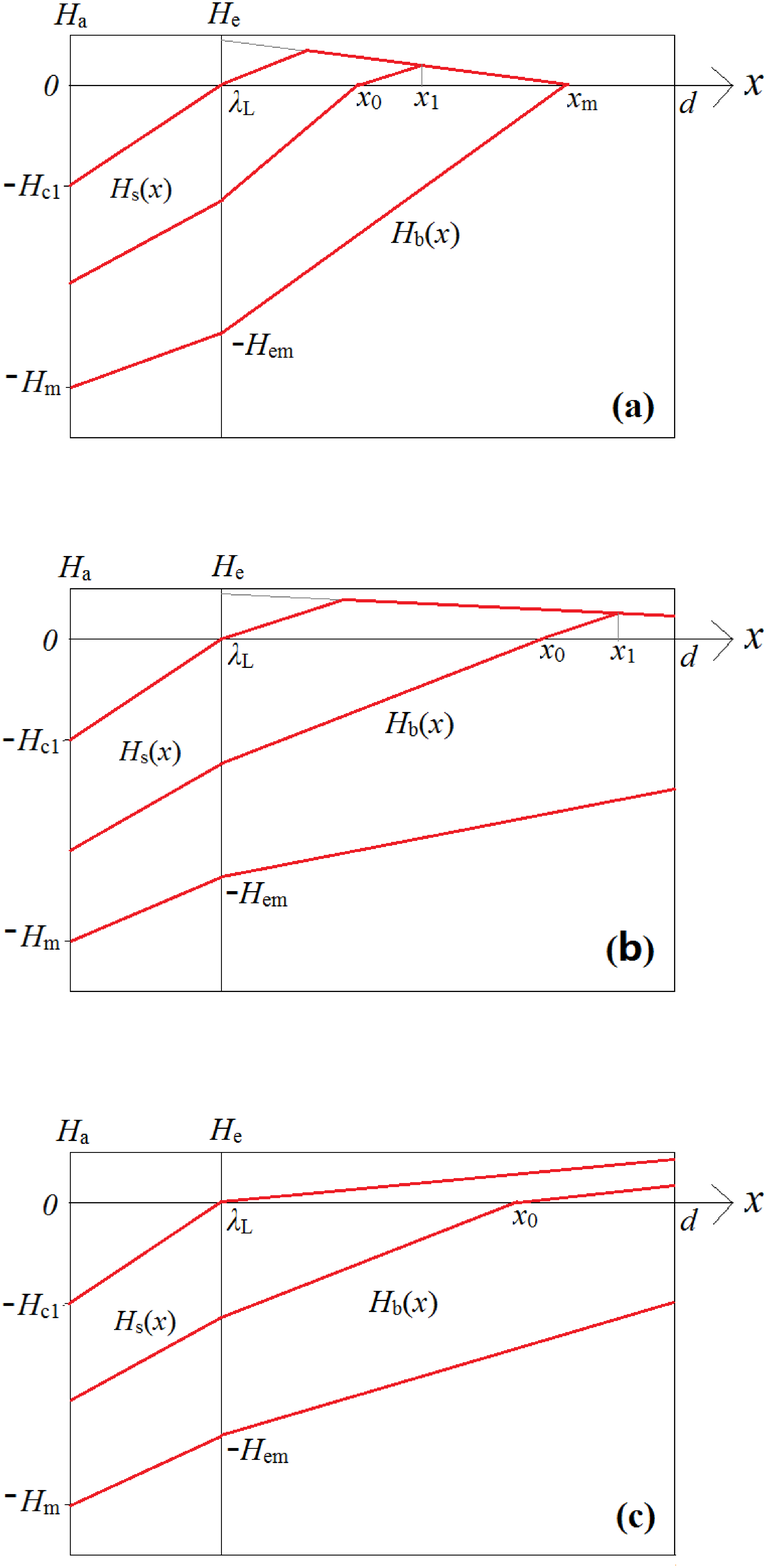}
\caption{Magnetic field distribution (red lines) of a superconductor in Stage ($\mathbf{v}$) ($-H_\mathrm{m} < H_\mathrm{a} < -H_\mathrm{c1}$) if: (a) $H_\mathrm{m} < H_\mathrm{fp}$; (b) $H_\mathrm{m} > H_\mathrm{fp}$ and $x_\mathrm{2} < d$; (c) $H_\mathrm{m} > H_\mathrm{fp}$ and $x_\mathrm{2} > d$. $H_\mathrm{s}(x)$ is the distribution function in the surface layer. $H_\mathrm{b}(x)$ is the distribution function in the bulk region.
} \label{fig6}
\end{figure}

Stage ($\mathbf{v}$): $H_\mathrm{a}$ is increased from $-H_\mathrm{c1}$ to $-H_\mathrm{m}$. The superconductor is in mixed state and vortices with opposite direction start to penetrate the bulk region. The frontier of vortices with opposite direction is defined as $x_\mathrm{0}$ and will move towards the center of superconductor. There should be annihilation of vortices with opposite vorticities at this moving frontier line. $H_\mathrm{e}$ is not zero but described by Eq.~\ref{equ1} and Eq.~\ref{equ2}. The bulk pinning current density $J_\mathrm{c}$ of the penetrating part is given by Eq.~\ref{equ10}. The inner remaining pinning current density $J_\mathrm{cr}$ is assumed to remain unchanged for simplicity $J_\mathrm{cr1} = J_\mathrm{cr}[-H_\mathrm{c1}]$. As in Stage ($\mathbf{iv}$), if $H_\mathrm{m} < H_\mathrm{fp}$, the position $x_\mathrm{0}$ is given by
\begin{equation}\label{equ54}
x_\mathrm{0}=-\frac{H_\mathrm{e}}{J_\mathrm{c}}+\lambda_\mathrm{L}.
\end{equation}
The position dividing currents with different directions $x_\mathrm{1}$ is
\begin{equation}\label{equ81}
x_\mathrm{1}=\frac{H_\mathrm{em}+J_\mathrm{cm}\lambda_\mathrm{L}+J_\mathrm{cr1}x_\mathrm{0}}{J_\mathrm{cm}+J_\mathrm{cr1}}.
\end{equation}
Thus, the field distribution can be written as (as shown in Fig.~\ref{fig6}(a))
\begin{equation}\label{equ55}
H_\mathrm{s}(x)=-\frac{x}{\lambda_\mathrm{L}}(H_\mathrm{a}-H_\mathrm{e})+H_\mathrm{a}.
\end{equation}
\begin{equation}\label{equ56}
H_\mathrm{b}(x)=
\begin{cases}
J_\mathrm{c}(x-\lambda_\mathrm{L})+H_\mathrm{e}& \text{$\lambda_\mathrm{L}<x<x_\mathrm{0}$}\\
J_\mathrm{cr1}(x-x_\mathrm{0})& \text{$x_\mathrm{0}<x<x_\mathrm{1}$}\\
-J_\mathrm{cm}(x-\lambda_\mathrm{L})+H_\mathrm{em}& \text{$x_\mathrm{1}<x<x_\mathrm{m}$}\\
0& \text{$x_\mathrm{m}<x<d$}
\end{cases}
\end{equation}
If $H_\mathrm{m} > H_\mathrm{fp}$ and $x_\mathrm{2} < d$, there are remaining pinning currents with different directions in the bulk region. $H_\mathrm{s}(x)$ is described by Eq.~\ref{equ55}. When $H_\mathrm{a} < H_\mathrm{fp}$, $H_\mathrm{b}(x)$ is (as shown in Fig.~\ref{fig6}(b))
\begin{equation}\label{equ58}
H_\mathrm{b}(x)=
\begin{cases}
J_\mathrm{c}(x-\lambda_\mathrm{L})+H_\mathrm{e}& \text{$\lambda_\mathrm{L}<x<x_\mathrm{0}$}\\
J_\mathrm{cr1}(x-x_\mathrm{0})& \text{$x_\mathrm{0}<x<x_\mathrm{1}$}\\
-J_\mathrm{cm}(x-\lambda_\mathrm{L})+H_\mathrm{em}& \text{$x_\mathrm{1}<x<d$}
\end{cases}
\end{equation}
But when $H_\mathrm{a} > H_\mathrm{fp}$, $H_\mathrm{b}(x)$ is (Fig.~\ref{fig6}(b))
\begin{equation}\label{equ60}
H_\mathrm{b}(x)=J_\mathrm{c}(x-\lambda_\mathrm{L})+H_\mathrm{e}.
\end{equation}
If $H_\mathrm{m} > H_\mathrm{fp}$ and $x_\mathrm{2} > d$, the remaining pinning current has only one direction. $H_\mathrm{s}(x)$ is described by Eq.~\ref{equ55}. When $H_\mathrm{a} < H_\mathrm{fp}$, $H_\mathrm{b}(x)$ is (as shown in Fig.~\ref{fig6}(c))
\begin{equation}\label{equ62}
H_\mathrm{b}(x)=
\begin{cases}
J_\mathrm{c}(x-\lambda_\mathrm{L})+H_\mathrm{e}& \text{$\lambda_\mathrm{L}<x<x_\mathrm{0}$}\\
J_\mathrm{cr1}(x-x_\mathrm{0})& \text{$x_\mathrm{0}<x<d$}
\end{cases}
\end{equation}
But when $H_\mathrm{a} > H_\mathrm{fp}$, $H_\mathrm{b}(x)$ is just as Eq.~\ref{equ60} (as shown in Fig.~\ref{fig6}(c)).

3. The field descending process in the negative field side

Stage ($\mathbf{vi}$): $H_\mathrm{a}$ is decreased from $-H_\mathrm{m}$ to $-H_\mathrm{c1}$. The superconductor is in mixed state and the penetrating process is the same as Stage ($\mathbf{iii}$), while both the magnetic field and the supercurrent have opposite direction.

Stage ($\mathbf{vii}$): $H_\mathrm{a}$ is decreased from $-H_\mathrm{c1}$ to 0 and then increased from 0 to $H_\mathrm{c1}$. The superconductor is in Meissner state and the penetrating process is the same as in Stage ($\mathbf{iv}$), while both the magnetic field and the supercurrent have opposite direction.

Stage ($\mathbf{viii}$): $H_\mathrm{a}$ is increased from $H_\mathrm{c1}$ to $H_\mathrm{m}$. The superconductor is in mixed state and the penetrating process is the same as Stage ($\mathbf{v}$), while both the magnetic field and the supercurrent have opposite direction.

\begin{figure}[htbp]
\centering
\includegraphics[width=8cm]{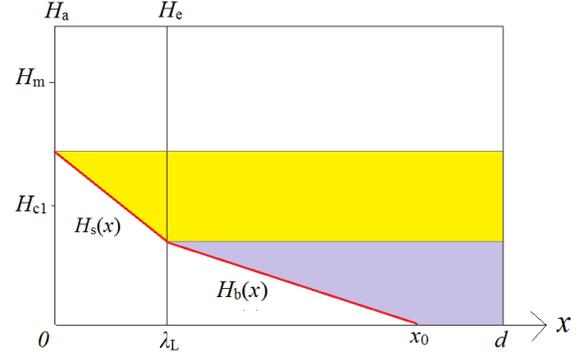}
\caption{Magnetic field distribution (red lines) of a superconductor in Stage ($\mathbf{ii}$). The yellow area denotes the equilibrium magnetization $M_\mathrm{equ}$, the purple area denotes the non-equilibrium magnetization $M_\mathrm{pin}$, respectively.
} \label{fig7}
\end{figure}

\begin{figure}[htbp]
\centering
\includegraphics[width=8cm]{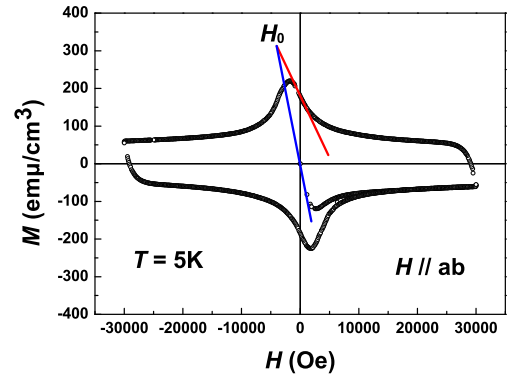}
\caption{An example showing how to determine the characteristic field $H_\mathrm{0}$, which is the field for the remaining bulk pinning current density $J_\mathrm{cr}$ to be zero. The black hallow points are the MHL of optimally doped Ba$_{0.6}$K$_{0.4}$Fe$_2$As$_2$ single crystal at 5K from our experiments. The blue line is the linear relation of $M_\mathrm{ini}$ (magnetization of the initial Meissner state), the red line is the linear relation of $M_\mathrm{des}$ (magnetization of the field descending Meissner state), and $H_\mathrm{0}$ is the intersection point of red line and blue line.
} \label{fig8}
\end{figure}

Based on the discussions above, the magnetization of superconducting plate can be calculated directly. Firstly, the total magnetization $M$ is divided into two parts, i.e., the equilibrium magnetization $M_\mathrm{equ}$ and the non-equilibrium magnetization $M_\mathrm{pin}$. For example, we take stage ($\mathbf{ii}$) in the initial penetrating process into account, and other stages can be obtained in the same way. The field distribution of stage ($\mathbf{ii}$) is shown in Fig.~\ref{fig7}. We mark different parts of magnetization with different colored areas. The yellow area denotes the equilibrium magnetization $M_\mathrm{equ}$, and the purple area denotes the non-equilibrium magnetization $M_\mathrm{pin}$, respectively. In this case, we have $M_\mathrm{equ}$ and $M_\mathrm{pin}$ given by
\begin{equation}\label{equ13}
M_\mathrm{equ} = -\frac{1}{d}[H_\mathrm{a}d-H_\mathrm{e}(d-\lambda_\mathrm{L})-\int_{0}^{\lambda_\mathrm{L}}H_\mathrm{s}(x)\mathrm{d}x]
\end{equation}
\begin{equation}\label{equ14}
M_\mathrm{pin} = -\frac{1}{d}[H_\mathrm{e}(d-\lambda_\mathrm{L})-\int_{\lambda_\mathrm{L}}^{d}H_\mathrm{b}(x)\mathrm{d}x]
\end{equation}
Adding together these two different terms, we have the total magnetization $M$:
\begin{equation}\label{equ15}
\begin{aligned}
M &= M_\mathrm{equ}+M_\mathrm{pin}\\
&= -\frac{1}{d}[H_\mathrm{a}d-\int_{0}^{\lambda_\mathrm{L}}H_\mathrm{s}(x)\mathrm{d}x-\int_{\lambda_\mathrm{L}}^{d}H_\mathrm{b}(x)\mathrm{d}x]
\end{aligned}
\end{equation}
With the field distribution $H_\mathrm{s}(x)$ in the surface layer and $H_\mathrm{b}(x)$ in the bulk region, the total magnetization can be calculated in this simple and effective way.

\begin{figure}[htbp]
\centering
\includegraphics[width=8cm]{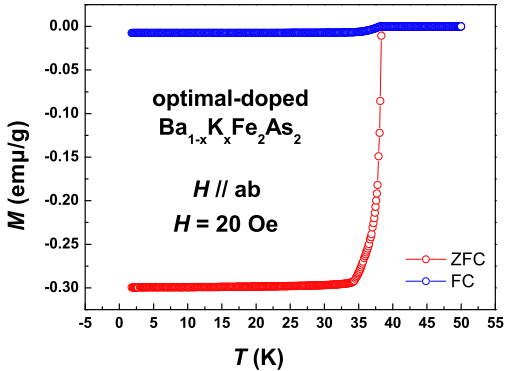}
\caption{Temperature dependence of magnetization measured with zero-field cooled (ZFC) and field cooled (FC) modes for the optimally doped Ba$_{0.6}$K$_{0.4}$Fe$_2$As$_2$ single crystal. The applied field is 20 Oe and parallel to the lateral $\mathbf{ab}$-plane.
} \label{fig9}
\end{figure}
In the end of this section, we will give the derivation of $H_\mathrm{0}$ in section II. According to our model, when $H_\mathrm{a}$ is reduced from a high value to below $H_\mathrm{c1}$, the superconductor will try to re-enter the Meissner state. In this case, the magnetization will exhibit a linear dependence on $H_\mathrm{a}$. But the slope of $\mathrm{d}M/\mathrm{d}H_\mathrm{a}$ is lower than that of the initial Meissner state. This is because the total magnetization of both the shielding part and the bulk pinning has a linear dependence on $H_\mathrm{a}$. Actually, the characteristic field $H_\mathrm{0}$, which defines the field for remaining pinning current density reduces to zero in our model, can be obtained from the experimental data based on this logic. It is given by the field associating with the intersection point of linear parts from magnetization curves in the initial Meissner state and the field descending Meissner state in the side of negative field. This can be understood in the following way. The total magnetization of the initial Meissner state is denoted as $M_\mathrm{ini}$ and given as
\begin{equation}\label{equ16}
M_\mathrm{ini} = -\frac{1}{d}[H_\mathrm{a}d-\int_{0}^{\lambda_\mathrm{L}}H_\mathrm{s}(x)\mathrm{d}x]= H_\mathrm{a}(\frac{\lambda_\mathrm{L}}{2d}-1)
\end{equation}
The derivative of $M_\mathrm{ini}$ with $H_\mathrm{a}$ is constant $\lambda_\mathrm{L}/2d-1$. In this case, the linear relation of $M_\mathrm{ini}$ in the vicinity of $H_\mathrm{a} = 0$ can be written as
\begin{equation}\label{equ17}
M_\mathrm{ini,0} = H_\mathrm{a}\frac{\mathrm{d}M_\mathrm{ini}}{\mathrm{d}H_\mathrm{a}} = H_\mathrm{a}(\frac{\lambda_\mathrm{L}}{2d}-1)
\end{equation}
On the other hand, the magnetization of the field descending Meissner state is denoted as $M_\mathrm{des}$. When $H_\mathrm{a} < H_\mathrm{c1}$, $M_\mathrm{des}$ is given as
\begin{equation}\label{equ18}
\begin{aligned}
M_\mathrm{des} &= -\frac{1}{d}[H_\mathrm{a}d-\int_{0}^{\lambda_\mathrm{L}}H_\mathrm{s}(x)\mathrm{d}x-\int_{\lambda_\mathrm{L}}^{d}H_\mathrm{b}(x)\mathrm{d}x]\\
&= H_\mathrm{a}(\frac{\lambda_\mathrm{L}}{2d}-1)+J_\mathrm{c0}\frac{H_\mathrm{a}-H_\mathrm{0}}{H_\mathrm{c1}-H_\mathrm{0}}\frac{(d-\lambda_\mathrm{L})^{2}}{2d}
\end{aligned}
\end{equation}
In the vicinity of $H_\mathrm{a} = 0$, we have $M_\mathrm{des}$
\begin{equation}\label{equ19}
M_\mathrm{des}\mid_{H_\mathrm{a}=0} = J_\mathrm{c0}\frac{-H_\mathrm{0}}{H_\mathrm{c1}-H_\mathrm{0}}\frac{(d-\lambda_\mathrm{L})^{2}}{2d}
\end{equation}
The derivative of $M_\mathrm{des}$ at the field $H_\mathrm{a} = 0$ is
\begin{equation}\label{equ20}
\frac{\mathrm{d}M_\mathrm{des}}{\mathrm{d}H_\mathrm{a}}\mid_{H_\mathrm{a}=0} = (\frac{\lambda_\mathrm{L}}{2d}-1)+\frac{J_\mathrm{c0}}{H_\mathrm{c1}-H_\mathrm{0}}\frac{(d-\lambda_\mathrm{L})^{2}}{2d}
\end{equation}
Thus, the linear relation of $M_\mathrm{des}$ in the vicinity of $H_\mathrm{a} = 0$ can be written as
\begin{equation}\label{equ21}
\begin{aligned}
M_\mathrm{des,0} &= H_\mathrm{a}\frac{\mathrm{d}M_\mathrm{des}}{\mathrm{d}H_\mathrm{a}}\mid_{H_\mathrm{a}=0}+M_\mathrm{des}\mid_{H_\mathrm{a}=0}\\
&= H_\mathrm{a}[(\frac{\lambda_\mathrm{L}}{2d}-1)+\frac{J_\mathrm{c0}}{H_\mathrm{c1}-H_\mathrm{0}}\frac{(d-\lambda_\mathrm{L})^{2}}{2d}]\\
&-\frac{J_\mathrm{c0}H_\mathrm{0}}{H_\mathrm{c1}-H_\mathrm{0}}\frac{(d-\lambda_\mathrm{L})^{2}}{2d}
\end{aligned}
\end{equation}
It can be derived that, the intersection of these two straight lines (let $M_\mathrm{ini,0} = M_\mathrm{des,0}$ from Eq.~\ref{equ17} and Eq.~\ref{equ21}) gives out $H_\mathrm{a} = H_\mathrm{0}$. In this way, the value of $H_\mathrm{0}$ is determined from experimental data, see an example in Fig.~\ref{fig8}. It can be seen that, the characteristic field $H_\mathrm{0}$ has a clear physical meaning and does not need to be treated as a fitting parameter.

To summarize for this section, we can simulate the flux penetrating process and fit the MHLs of type-II superconductors well by using proper fitting parameters. This plays an important role in studying the physical properties of type-II superconductors.

\begin{figure*}[htbp]
\centering
\includegraphics[width=16cm]{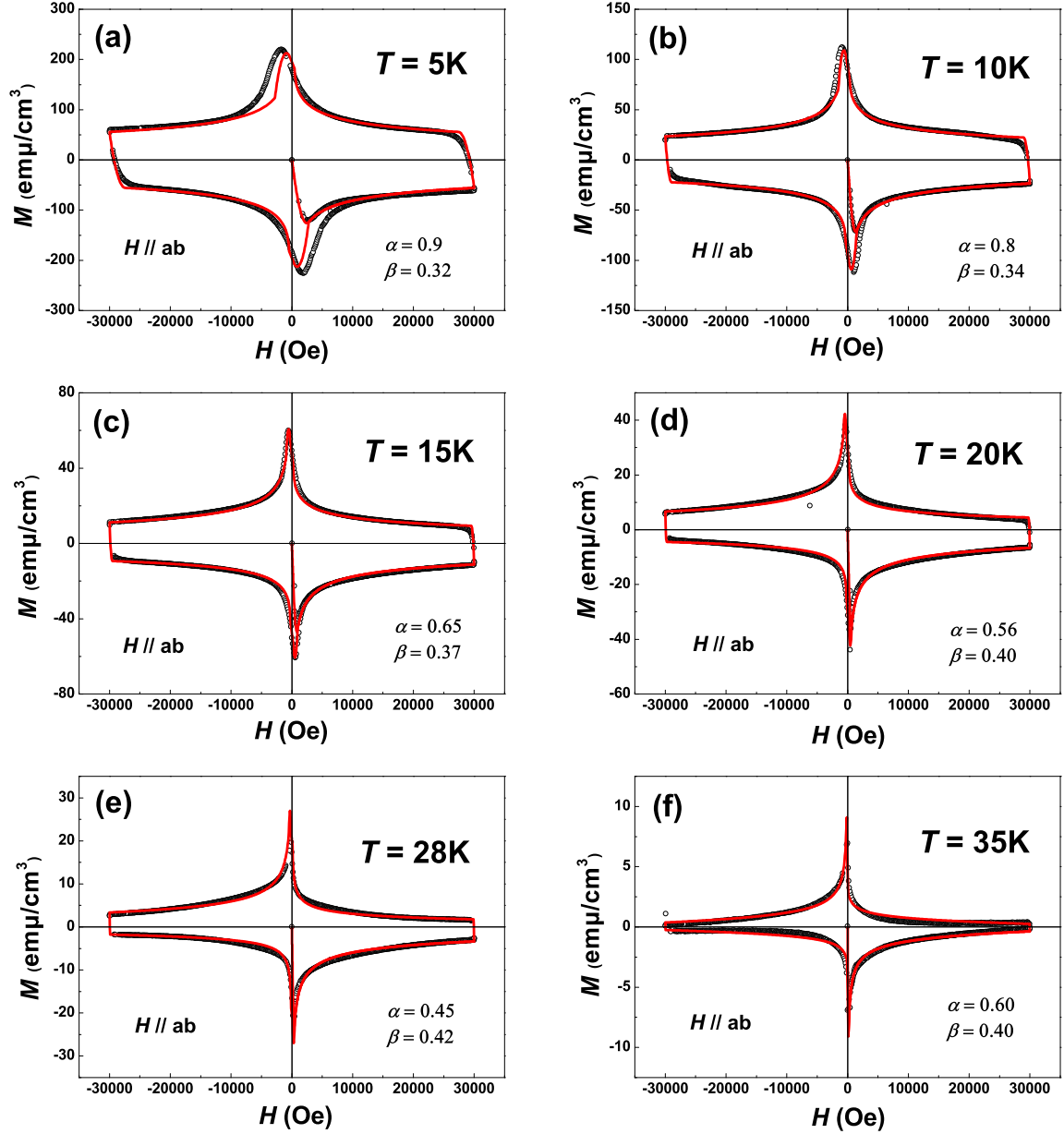}
\caption{MHLs (black hallow points) and corresponding fitting curves (red lines) of the optimally doped Ba$_{0.6}$K$_{0.4}$Fe$_2$As$_2$ single crystal, at temperatures of: (a) 5K, (b) 10K, (c) 15K, (d) 20K, (e) 28K, (f) 35K. For the MHL of 35K, a small background magnetization due to the induced magnetic field of the measuring coils has been subtracted, which was measured with no sample on the sample holder. This background magnetization is too small and can be ignored for other temperatures.
} \label{fig10}
\end{figure*}
\section{Experiments and Model Fits}
To check how effective of our model is, the MHLs of high-quality iron-based superconductors are studied. The sample used in the experiments is a Ba$_{0.6}$K$_{0.4}$Fe$_2$As$_2$ single crystal, grown by self-flux method using FeAs as flux \cite{BaK122}. The dimensions of the sample are 1.8mm$\times$1.1mm$\times$0.15mm and the weight is 1.70mg. The DC magnetization measurements were carried out on a SQUID-VSM-7T (Quantum Design). The applied field was parallel to the lateral $\mathbf{ab}$-plane. The field was swept from 0 to 3T, then to -3T and finally to 3T again with a rate of 50Oe/s . The temperature dependence of magnetization $M(T)$ is shown in Fig.~\ref{fig9} and the sharp transition indicates that the sample is of high quality. The obtained critical temperature $T_\mathrm{c}$ is 38.5K, which indicates that the sample is close to the optimal doping point \cite{BaK122}. We need to note that, the small tail on the transition curve makes the transition a little broad, but we find that this feature is intrinsic, since the transition curve when the field is perpendicular to the $\mathbf{ab}$-plane looks very sharp. This tail may be attributed to a divergent London penetration depth when temperature is approaching $T_\mathrm{c}$.

\begin{figure*}[htbp]
\centering
\includegraphics[width=13cm]{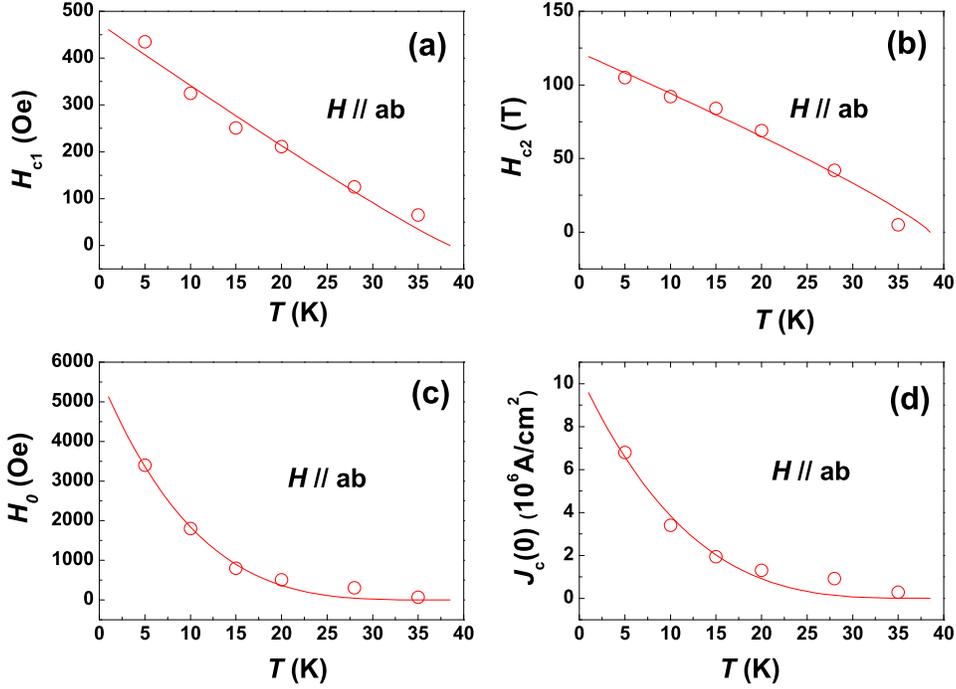}
\caption{Temperature dependence of different fitting parameters. (a) The lower critical field $H_\mathrm{c1}$; (b) The upper critical field $H_\mathrm{c2}$; (c) The characteristic field $H_\mathrm{0}$ for remaining bulk pinning current density reduced to zero; (d) The zero-field critical current density $J_\mathrm{c0}$. The red circles are fitting parameters at different temperatures. The red lines are empirical scaling relations.
} \label{fig11}
\end{figure*}
The MHLs of the optimally doped Ba$_{0.6}$K$_{0.4}$Fe$_2$As$_2$ single crystal were measured at temperatures of 5K, 10K, 15K, 20K, 28K, 35K. We fit all the MHLs with our model. The experimental data and the corresponding fitting curves are shown in Fig.~\ref{fig10}, respectively. All data shown here are the raw data beside that for 35K, in which a small background magnetization due to a remanent magnetic field of the superconducting coils is subtracted, which was measured without sample on the sample holder. This background magnetization is too small and can be ignored for other temperatures. The London penetration depth used in the fitting process is $\lambda_\mathrm{L}$ = 300nm from MFM measurements as determined for the similar sample \cite{depth}. We assume the London penetration depth to be independent of temperature for simplicity. This is a good approximation as the London penetration depth $\lambda_\mathrm{L}$ is much lower than the thickness of our sample $2d$ ($2d$ = 0.15mm, $\lambda_\mathrm{L} \ll 2d$). It can be seen in Fig.~\ref{fig10} that, the fitting curves accord well with the experimental data for almost all temperatures, but deviates slightly in the low-field region at 5K in the field penetration process. This may be attributed to the surface Bean$-$Livingston barrier which is not considered in our model \cite{surface,BL,Majer,Zeldov}. The Bean$-$Livingston barrier can prevent the vortices from entering the superconductor and may play an important role especially when the applied field is low. This effect becomes weaker when the temperature is increased to $T_\mathrm{c}$. In addition, we find that MHL becomes more asymmetric when the measuring temperature becomes higher. This indicates that, the ratio of equilibrium magnetization in the total magnetization rises with increasing temperature \cite{surface,Gokhfeld1,Gokhfeld2}. On the other hand, the non-equilibrium magnetization plays the leading role at temperatures lower than 10K, which leads to the symmetric MHLs with respect to the horizontal axis \cite{BeanPRL,BeanRMP}.

The temperature dependence of fitting parameters obtained from the fitting process are presented in Fig.~\ref{fig11}. The lower critical field $H_\mathrm{c1}(T)$ is shown in Fig.~\ref{fig11}(a), the upper critical field $H_\mathrm{c2}(T)$ is shown in Fig.~\ref{fig11}(b) and the characteristic field $H_\mathrm{0}(T)$ for the remaining bulk pinning current to be zero is shown in Fig.~\ref{fig11}(c), respectively. All three characteristic fields are dropping down with increasing temperature. An empirical scaling relation $H(T) = H(0)(1-T/T_\mathrm{c})^{n}$ was used to fit these three fields. In this case, we have obtained the following results, $H_\mathrm{c1}(0)$ = 474Oe and $n$ = 1.09 for $H_\mathrm{c1}(T)$, $H_\mathrm{c2}(0)$ = 122T and $n$ = 0.86 for $H_\mathrm{c2}(T)$, $H_\mathrm{0}(0)$ = 5655Oe and $n$ = 3.73 for $H_\mathrm{0}(T)$, respectively. The obtained value of $H_\mathrm{c1}(0)$ is a little lower than that given by Ren \cite{Hc1}, which was about 600Oe and determined simply from the deviation of Meissner lines. The value of $H_\mathrm{c2}(0)$ is lower than that predicted by Wang (235 T) \cite{Hc2} and Ishida (300 T) \cite{Ishida}, but higher than that given by Altarawneh et al. (75 T) \cite{anisotropyHc2}. Unfortunately there are no direct measurements about the $H_\mathrm{c2}(0)$ since it is really too high. However, both the magnitude and trend of $H_\mathrm{c2}$ obtained in our work in the high temperature region is consistent with those in reference \cite{Gurevich,Tarantini}. Fig.~\ref{fig11}(d) shows the temperature dependence of the zero-field critical current density $J_\mathrm{c0}(T)$. It can be seen that $J_\mathrm{c0}(T)$ decreases to zero when the measuring temperature is increased to $T_\mathrm{c}$. An empirical scaling relation $J(T) = J(0)(1-T/T_\mathrm{c})^{n}$  was also used to fit the data and determining the value of $J_\mathrm{c0}(0)$. The zero-field critical current density at 0K determined here is $J_\mathrm{c0}(0)$ = 1.046$\times$10$^{7}$ A/cm$^{2}$, higher than those given by others \cite{Ishida,vortexdynamics} which were obtained directly through the original Bean critical state model. The fitting exponent of $J_\mathrm{c0}(T)$ is $n$ = 3.34, close to $n$ = 3.73 of $H_\mathrm{0}(T)$, indicating a close relationship between these two parameters.

From above analysis and fitting to the experimental data, we find that our model which counts the different magnetizations from the surface layer and the bulk pinning region is quite effective. Although the model is quite simple in form, it can capture the fundamental physics of both regions with different contributions of magnetization. By adjusting fitting parameters and proportion of two parts of magnetization, our model can be used to describe the flux penetration and magnetization of various type-II superconductors with different values of $\kappa$ or ratios of $H_\mathrm{c2}/H_\mathrm{c1}$. For further verification of our model, more experiments of different superconductors are needed. In addition, the difference between simulated curve and experimental data near $H_\mathrm{c1}$ in low temperature region may be induced by the Bean-Livingston surface barrier \cite{BL,Majer,Zeldov}, on which further experiments are desired. The enhancement of magnetization due to this effect occurs mainly in the flux entry process, as shown by our experimental data, but unfortunately it cannot be explicitly expressed so far. Our model may be extended to other type-II superconductors when a global description on MHLs is needed.

\section{Conclusion}

In summary, a generalized phenomenological model for critical state has been proposed to describe the magnetic penetration and MHLs of type-II superconductors. The model combines the equilibrium magnetization of surface screening current and the non-equilibrium magnetization of bulk pinning, and deals with the vortex-trapped Meissner state in a more reasonable way. We use the model to simulate the MHLs of optimally doped Ba$_{0.6}$K$_{0.4}$Fe$_2$As$_2$ single crystal and the experimental data can be fitted quite well. Furthermore, our model can serve as an effective tool to study the magnetization hysteresis and vortex penetration of type-II superconductors with different values of $\kappa$ or ratios of $H_\mathrm{c2}/H_\mathrm{c1}$.

\begin{acknowledgments}
This work was supported by the National Natural Science Foundation of China (Grants: A0402/13001167, A0402/11534005 and A0402/11674164) and National Key R and D Program of China (Grant nos. 2016YFA0300401 and 2016YFA0401704), and the Strategic Priority Research Program of Chinese Academy of Sciences (Grant No.XDB25000000).
\end{acknowledgments}

$^\dag$ hhwen@nju.edu.cn

\end{document}